\documentclass[11pt]{article}
\usepackage{cospar}
\usepackage{url}

\usepackage{graphicx}
\usepackage[figuresright]{rotating}

\title{RELATIVISTIC ASTROPHYSICS EXPLORER}

\author{P. Kaaret\address{Harvard-Smithsonian Center for Astrophysics, 
    60 Garden St., Cambridge, MA 02138, USA}}

\begin{document}

\maketitle

\begin{abstract}

The great success of the Rossi X-Ray Timing Explorer (RXTE) has shown
that X-ray timing is an excellent tool for the study of strong
gravitational fields and the measurement of fundamental physical
properties of black holes and neutron stars.  Here, we describe a
next-generation X-ray timing mission, the Relativistic Astrophysics
Explorer (RAE), designed to fit within the envelope of a medium-sized
mission. The instruments will be a narrow-field X-ray detector array
with an area of 6~m$^2$ equal to ten times that of RXTE and a
wide-field X-ray monitor.  We describe the science made possible with
this mission, the design of the instruments, and results on prototype
large-area X-ray detectors.

\end{abstract}

\section*{INTRODUCTION}

Timing is a key tool of X-ray astronomy.  The first definitive source
identifications made using X-ray data alone were of X-ray pulsars --
identified via periodic signals with sinusoidal period modulation due
to orbital motion (Tananbaum et al.\ 1972).  Recently, X-ray timing has
made substantial and unique contributions to our understanding of
accreting compact objects including the behavior of matter in strong
gravitational fields, the formation of relativistic jets, the physical
geometry and emission mechanisms of active galactic nuclei, the
evolution of neutron stars in binaries, and X-ray emission regions in
cataclysmic variables by exploiting the large effective area of the
Rossi X-Ray Timing Explorer (RXTE) (for a review see Bradt 1999).

These successes are a strong indication that further advances in X-ray
timing instrumentation will engender further scientific advances.  To
take an `order of magnitude' step, a next generation X-ray timing
mission will need an X-ray detector with an area ten times that of
RXTE, on the order of 6~m$^2$ (Kaaret et al.\ 2001).  With an order of
magnitude increase in photon counting rate, it will be possible to go
beyond the initial steps made with RXTE and exploit high frequency
X-ray timing to test models of the behavior of matter in strong
gravitational fields and to measure the fundamental physical properties
of black holes and neutron stars.  In addition, the increased
sensitivity will make possible new discoveries of unanticipated
phenomena.  Here, we describe two examples highlighting application of
X-ray timing to the study of strong-field gravity and ultradense
matter.

\begin{figure} \centerline{\includegraphics[width=85mm]{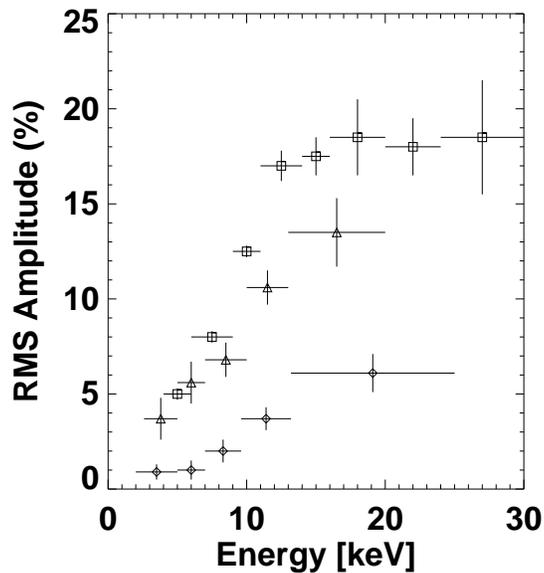}}
\caption{RMS amplitude versus energy for several high frequency signals
detected with RXTE.  The diamonds are for the 67~Hz QPO from the black
hole candidate GRS 1915+105 from Morgan, Remillard, and Greiner
(1997).  The squares are for the kHz QPO from the neutron star system
4U 1608-52 from Berger et al.\ (1996).  The triangles are for an X-ray
burst from KS 1731-260 from Muno et al.\ (2000).  In all cases,the rms
amplitude increases strongly at high X-ray energies above 10~keV.}
\end{figure}%

\subsection{Fast quasiperiodic oscillations from black hole
candidates} 

Quasiperiodic oscillations (QPOs)  have been discovered in the X-ray
emission from several accreting black hole candidate (BHC) systems with
frequencies up to 450~Hz. In two sources the fast QPO frequencies
appear constant regardless of the source state, while in a third the
QPO frequency varies by 10\% (Remillard et al.\ 1999).  The fast QPO
frequencies appear to be harmonically related, occurring at integer
multiples of a (usually absent) fundamental frequency.  A number of
models of the QPOs have been proposed, most of which involve
strong-field general relativistic effects.  One very intriguing model
explains the harmonic structure as due to resonances between orbital
and epicyclic motion around a rotating black hole (Abramowicz and
Kluzniak 2001). These fast QPOs from BHCs are weak (amplitudes near
1\%), often detected only at hard X-ray energies (above 12~keV), and
difficult to study in detail with RXTE.  Distinguishing amongst the
various models will be difficult with the current data or any data
likely to be obtained in the remaining lifetime of RXTE.  The increase
in the photon statistics with a 6~m$^2$ X-ray detector would  enable
reliable detection of the fast QPOs with relatively short integrations,
allowing accurate measurement of the QPO parameters and their
variations with time or correlations with spectral or other timing
parameters.  Such detailed measurements should lead to unique
identification of the physical process generating the QPOs. With a
unique model, it should be possible to exploit these QPOs to probe
strong-field gravity and measure the spin, and possibly mass, of the
black holes.

\subsection{Millisecond oscillations in X-ray bursts} 

Millisecond oscillations have been discovered in X-ray bursts from
several accreting neutron stars.  The detection of oscillations with
very high coherence in a long-duration X-ray burst (Strohmayer and
Markwardt 2002) and the detection of burst oscillations at the known
spin frequency of the millisecond pulsar SAX 1808.4-3658 (D.\
Chakrabarty 2002) show that the oscillations are related to the neutron
star spin and are likely due to inhomogeneous nuclear burning on a
rotating neutron star.  The burst oscillations probe conditions at the
neutron star surface and detailed measurements of their properties
should lead to constraints on the mass-radius relation of neutron stars
and therefore on the equation of state of ultradense matter.  Because
the strength of relativistic light bending due to the gravity of the
neutron star depends on the neutron-star mass-radius relation, the
observed maximum modulation in the burst oscillations places direct
constraints on the mass-radius relation (Strohmayer et al.\ 1998).   In
addition, the spectrum of the emitted radiation should be modified by
Doppler shifts as the line-of-sight velocity of the nuclear burning hot
spot changes as the neutron star rotates (Ford 1999).  This provides a
direct measure of the surface velocity and, hence, the neutron star
radius since the spin period is known.  A 6~m$^2$ X-ray detector would
detect roughly 1000 counts in each oscillation cycle near the peak of a
bright X-ray burst.  This would permit detailed examination of
individual oscillation cycles and allow accurate measurement of the
modulation amplitude and Doppler shifts  for individual oscillation
cycles.

\begin{figure}  \centerline{\includegraphics[width=150mm]{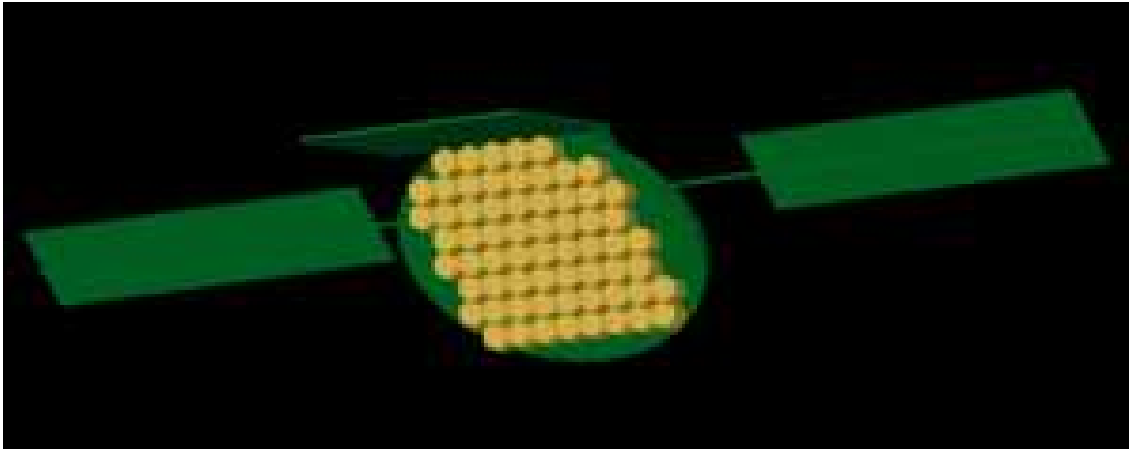}}
\caption{Perspective view of the Relativistic Astrophysics Explorer --
a satellite for X-ray timing with a modular, large-area X-ray
detector.} \end{figure}%

\section*{INSTRUMENT REQUIREMENTS}

The key characteristic of RXTE is its large effective area.  To achieve
an order of magnitude increase in X-ray timing capabilities, a ``next
generation'' X-ray timing mission will need a  large area X-ray
detector with  an effective area equal to ten times that of RXTE, on
the order of 6~m$^2$ (Kaaret et al.\ 2001; Barret et al.\ 2001).   Good
sensitivity extending up to energies of 20 to 30~keV is required
because the modulation of timing signals detected from Galactic
accreting objects is, in general, much higher at high energies. 
Figure~1 shows the rms modulation versus energy for several typical
high frequency signals detected with RXTE.  The strong increase at high
energies is clear.  Because the time required for detection of a signal
at fixed source counting rate scales inversely with the fourth power of
the rms modulation and the detection significance for a fixed
integration time scales as the square of the rms modulation, there are
great advantages to be gained from a timing instrument with significant
collecting area at high energies.  Furthermore, there are some signals,
such as the higher frequency QPOs from accreting black holes candidates
discussed in the previous section that are detected only at high
energies.  For GRO J1655-40, the second QPO at higher frequency was
detected only at energies above 13~keV (Strohmayer 2001).

Focusing of X-rays in the hard band is difficult because of the
extremely small graze angles required for efficient broad-band
reflection and it is unlikely that focusing telescopes will achieve the
required areas, of order 6~m$^2$, in the near future.  The Hard X-ray
Telescope (HXT) on the Constellation-X mission is expected to have a
total area of less than 0.6~m$^2$ at 10~keV, and even the extended XEUS
mission may achieve only 1.7~m$^2$ at 10~keV with the area decreasing
rapidly at higher energies (Aschenbach et al.\ 2001).  Integrated over
the band where the second QPO from GRO J1655-40 was detected, the
effective area of XEUS is less than that of the RXTE PCA.  While some
advance in X-ray timing will be possible with the currently planned
major observatories, a significant increase in X-ray timing 
capabilities will likely require a dedicated mission.  Only moderate
energy resolution ($\sim 1 \rm \, keV$ at 6~keV) and no position
resolution are required.  The detector may be non-imaging with a field
of view limited by a collimator.  Better energy resolution is
desirable, but is not essential for X-ray timing.

Non-focusing instruments do not suffer from the difficulties of
reflecting hard X-ray photons and large effective area in the
10--30~keV band is relatively easily achieved.  The high energy
response is typically limited by absorption depth in the detector. 
High efficiency up to 20 or 30~keV can be achieved by gaseous or solid
state detectors as described in the next section.   The primary
disadvantage of a non-focusing instrument is the relatively high
background counting rate.  For this reason, great care must be taken to
actively veto background events.  Large non-focusing instruments can
easily be constructed in a modular fashion.  In addition to reducing
costs and facilitating construction, dividing the full effective area of
the instrument into multiple independent units decreases the count rate
for each unit.  This is particularly advantageous for a timing mission
because the effect of instrumental dead time decreases, making it
possible to achieve very high total count rates with minimal dead time
effects.

A large detector area with sensitivity over a broad band extending  up
to high energies (2-30~keV) is the critical requirement for achieving
the very high source counting rate essential for the success of a
future X-ray timing mission.  To effectively exploit such high rates, a
high telemetry bandwidth is also required.  To fully telemeter the
event rate of $2 \times 10^5$~c/s from a bright source, comparable in
intensity to the Crab nebula, will require a telemetry rate of the 
order of 10~Mbps.  Multiple ground stations may be needed to achieve
the required average telemetry rate. To buffer events while out of
ground contact and for very bright sources, a large on-board memory (of
order 100~Gbit) will be required.

Most of the sources of interest for high frequency X-ray timing are
variable or transient.  To maximize the scientific return of a timing
mission, or timing observations in general, it is critical to have an
all-sky monitor sufficiently sensitive to discover and localize new
transients and determine the state of known persistent sources. In
addition, flexible observation scheduling and rapid response to targets
of opportunity are needed in order to perform observations while the
sources are in interesting states.  This is something that may be very
difficult to achieve with a multi-purpose large observatory and is a
strong advantage for a dedicated mission.  An X-ray timing mission
should be designed so that a large fraction of the sky is accessible at
any instant and so that frequent repointing can be performed in order
to contemporaneously monitor multiple targets.

\begin{figure}  \centerline{\includegraphics[width=75mm]{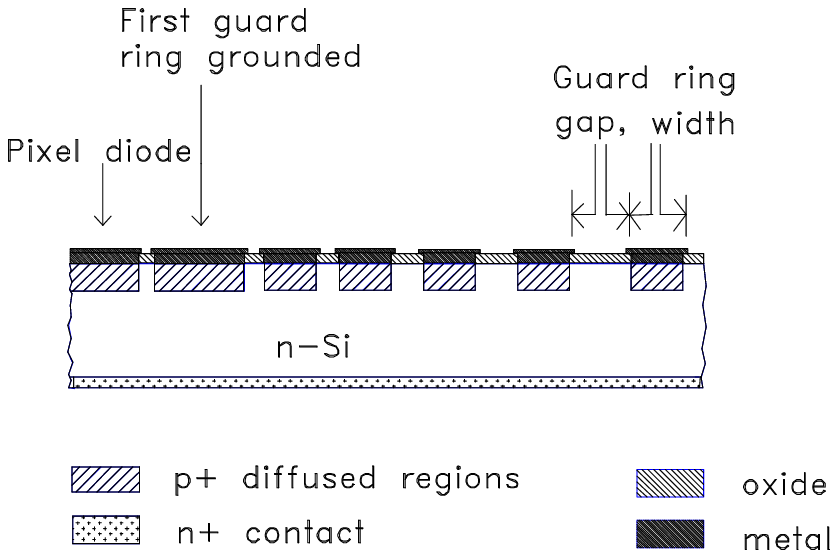} ~
\includegraphics[width=75mm]{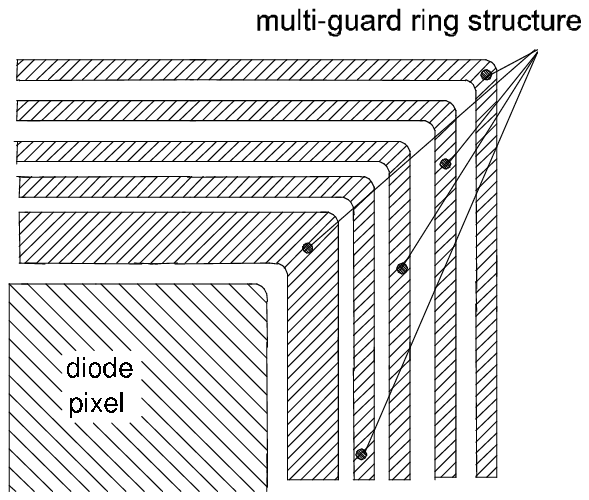}} \caption{Schematics of a
detector with generalized multiguard structure: cross-section of pixel
edge (left) and top view of pixel corner (right).} \end{figure}%

\section*{DETECTOR TECHNOLOGIES}

The primary technology driver for a new X-ray timing mission will be
the large area X-ray instrument.  We have considered both gas
proportional counters and solid-state silicon detectors for the
detection technology.  Gas proportional counters have a long history in
space.  However, proportional counters require delicate construction,
need high-voltage, and are notoriously tricky to build and finicky to
operate, particularly after the first one or two years of operation.
Each proportional counter unit (PCU) in the PCA on RXTE currently
requires attention to its history of operation with each PCU being
cycled on and off to prevent sparking and detector damage.  More
advanced proportional counter technologies have been developed
recently, including gas electron multipliers and microwell arrays, but
these are lacking in flight heritage.

Silicon is, perhaps, the most widely used radiation detection medium
today.  Silicon has a long heritage in radiation detection
applications, including specifically satellite-borne X-ray detectors
which have been operated successfully in  high radiation environments
(e.g.\ Aukerman, Vernon, \& Song 1984).  Silicon detectors are operated
without high-voltage or internal amplification, and are robust and have
stable operation over long periods.  Furthermore, the technology for
processing silicon is well understood and widely available.  Hence,
detectors, even those requiring large amounts of material, can be
fabricated by industry with existing facilities and at reasonable cost.

The simplest silicon detector is the p-i-n photodiode, and we have
studied the suitability of p-i-n photodiode detectors for an large-area
X-ray detector.  The energy resolution of a large area p-i-n photodiode
detector is limited by electronic noise originating in the device. 
Good energy resolution, better than $\Delta E = 1$~keV at 6~keV, can be
achieved for devices with areas of 1~cm$^2$. Assuming an area for each
individual detector pixel of $0.6 \rm \, cm^{2}$, a total of $10^{5}$
pixels would be required for a $6 \rm \, m^{2}$ detector.  One
electronics channel, with preamplifier, shaping amplifier, and
sample/hold, would then be required for each pixel, leading to a total
electronics channel count of $10^{5}$.  With an expected power per
channel near 2~mW, the total power required is near 200 W.  Our
preliminary studies and also those done for the European Experiment for
X-ray Timing and Relativistic Astrophysics (EXTRA) proposal (Barret
2001) show that p-i-n photodiode detectors can meet the scientific
requirements of an X-ray timing mission.

Until now, silicon detectors have been limited primarily to the soft or
standard X-ray band (below 10~keV) due to the limited absorbing power
of the relatively thin (typically $300 \mu$m or less) wafers commonly
used by the microelectronics industry.  To extend the reach of
silicon-based detectors to higher energies, thick devices must be used
to obtain sufficient stopping power.  The key problem with thick
silicon detectors is that a high bias voltage must be applied to fully
deplete the silicon and allow efficient collection of the charge
produced by incident X-rays.  Many previous attempts to fabricate thick
silicon detectors have failed because either the detectors break down
when the bias voltage is applied or the bias required causes large dark
currents, leading to poor performance. Recently, some success has been
obtained by Ota et al.\ (1999) who developed thick p-i-n diodes for the
hard X-ray instrument on Astro-E (now Astro-E2) and by Phlips et al.\
(2001) who demonstrated the X-ray detection capabilities of a thick
diode developed for heavy-ion detection.

Multi-guard ring structures allow high bias voltages to be applied
without causing excessive dark currents or premature detector breakdown
(Evensen et al.\ 1993; Avset \& Evensen 1996; Da Rold et al.\ 1997) . 
Figure~3 shows schematics of such devices.  Working with Photon
Imaging, Inc., we have designed new multi-guard ring structures and
fabricated devices from thick silicon wafers with good efficiency up to
30~keV.  The devices were fabricated by Photon Imaging from high
resistivity silicon to lower the depletion voltage.  The starting
material was carefully selected and the pixel p/n junction diffusion
structures were optimized  to further increase the breakdown voltage
(Sze \& Gibbons 1966; Waver 1972; Beck et al.\ 1996).  These advanced
designs produced very low dark currents, leading to 1.5~mm thick
detectors with good X-ray performance.

\begin{figure}  \centerline{\includegraphics[width=75mm]{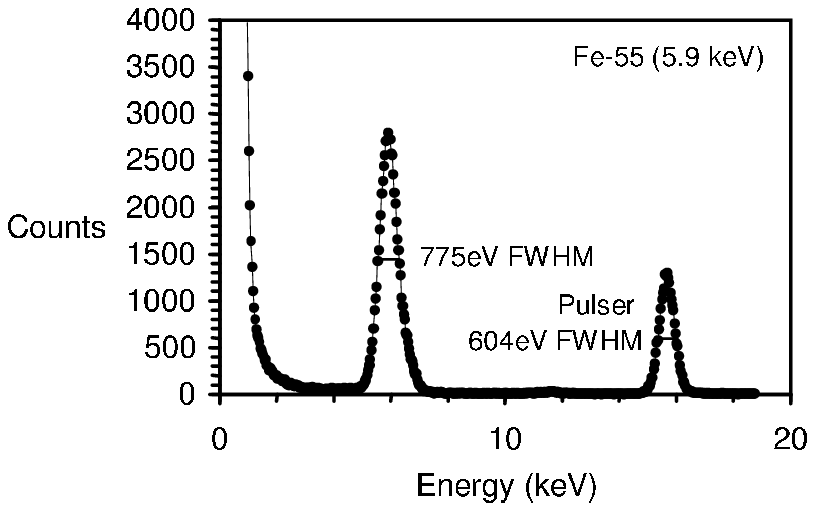} ~
\includegraphics[width=75mm]{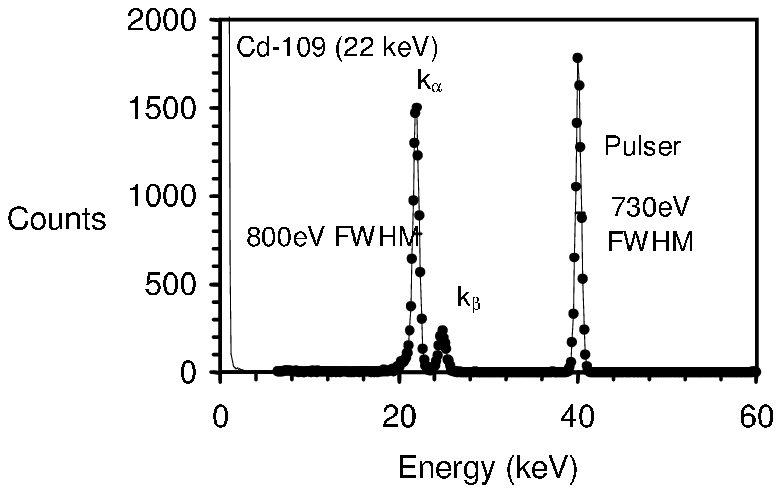}} \caption{Response of a 1.5~mm
thick, 25~mm$^2$ diode to various radioactive sources.} \end{figure}%

Figure~4 shows the spectral response of 1.5~mm thick diodes to X-rays
emitted by the radioactive isotopes $^{55}$Fe and $^{109}$Cd.  The
X-ray response was measured with the diodes fully depleted at 500~V.  A
three-stage thermoelectric cooler was used for cooling to $-30^{\circ}
\rm \, C$.  A standard low noise NIM-based amplifier and a PC-based
multi-channel analyzer were used for data collection.  The preamplifier
was not optimized for the leakage currents and capacitances of such
large diodes.  For the 1.5~mm thick diodes with an area 25~mm$^2$
cooled to $-30^{\circ} \rm \, C$, the measured energy resolution is
775~eV FWHM at 5.9~keV and 800~eV at 22~keV.  This energy resolution
meets the scientific requirements for a new X-ray timing mission.

\section*{CONCLUSION}

There have been only two US-lead major X-ray astrophysics observatories
in the past decade: RXTE and Chandra.  Although Chandra was very
recently launched, next-generation capabilities in X-ray imaging and
spectroscopy are already under serious development. RXTE has addressed
scientific issues, including relativistic gravity and the physics of
extreme phenomena, which are central to NASA's scientific goals. 
Further advances in X-ray timing, which are unlikely to come from any
currently planned mission but are possible with a dedicated mission,
will likely lead to further significant advances in our understanding
of compact objects and accretion in strong gravitational fields.


\end{document}